\documentclass[12pt,preprint]{aastex}

\usepackage{amsmath}
\usepackage{graphicx}
\usepackage{comment}
\usepackage{color} 
\newcommand{\fig}[1]{Fig.~\ref{#1}}
\newcommand{\tbl}[1]{Table~\ref{#1}}
\newcommand{\sect}[1]{Section~\ref{#1}}

\begin{document}
\title{Influence of photospheric magnetic conditions on the catastrophic behaviors of flux ropes in active regions}
\author{Quanhao Zhang\altaffilmark{1}, Yuming Wang\altaffilmark{1,2}, Youqiu Hu\altaffilmark{1}, Rui Liu\altaffilmark{1,3}, Jiajia Liu\altaffilmark{1,4}}
\altaffiltext{1}{CAS Key Laboratory of Geospace Environment, Department of Geophysics and Planetary Sciences, University of Science and Technology of China, Hefei 230026, China}
\altaffiltext{2}{Synergetic Innovation Center of Quantum Information \& Quantum Physics, University of Science and Technology of China, Hefei, Anhui 230026, China}
\altaffiltext{3}{Collaborative Innovation Center of Astronautical Science and Technology, China}
\altaffiltext{4}{Mengcheng National Geophysical Observatory, School of Earth and Space Sciences, University of Science and Technology of China, Hefei 230026, China}
\email{zhangqh@mail.ustc.edu.cn}

\begin{abstract}
Since only the magnetic conditions at the photosphere can be routinely observed in current observations, it is of great significance to find out the influences of photospheric magnetic conditions on solar eruptive activities. Previous studies about catastrophe indicated that the magnetic system consisting of a flux rope in a partially open bipolar field is subject to catastrophe, but not if the bipolar field is completely closed under the same specified photospheric conditions. In order to investigate the influence of the photospheric magnetic conditions on the catastrophic behavior of this system, we expand upon the 2.5 dimensional ideal magnetohydrodynamic (MHD) model in Cartesian coordinates to simulate the evolution of the equilibrium states of the system under different photospheric flux distributions. Our simulation results reveal that a catastrophe occurs only when the photospheric flux is not concentrated too much toward the polarity inversion line and the source regions of the bipolar field are not too weak; otherwise no catastrophe occurs. As a result, under certain photospheric conditions, a catastrophe could take place in a completely closed configuration whereas it ceases to exist in a partially open configuration. This indicates that whether the background field is completely closed or partially open is not the only necessary condition for the existence of catastrophe, and that the photospheric conditions also play a crucial role in the catastrophic behavior of the flux rope system.
\end{abstract}

\keywords{Sun: filaments, prominences---Sun: coronal mass ejections (CMEs)---Sun: flares---Sun: magnetic fields}

\section{Introduction}
\label{sec:introduction}
Large-scale solar explosive phenomena, such as prominence/filament eruptions, flares and coronal mass ejections (CMEs), are widely considered to be different manifestations of the same physical process \citep[e.g.][]{Low1996a,Archontis2008a,Chen2011a,Zhang2014a}, which is believed to be closely related to solar magnetic flux ropes \citep[e.g.][]{Low2001a,Torok2011a}. Many theoretical analyses have been made to investigate the eruptive mechanisms of magnetic flux ropes so as to shed light on the physical processes of solar eruptive activities \citep{Forbes1995a,Chen2000a,Kliem2006a,Su2011a,Longcope2014a}. \cite{vanTend1978a} concluded that a filament system loses equilibrium if the current in the filament exceeds a critical value. This process is called ``catastrophe'', which occurs via a catastrophic loss of equilibrium. Catastrophe has been suggested to be responsible for flux rope eruptions by many authors \citep{Priest1990a,Forbes1991a,Isenberg1993a,Lin2004b,Zhang2007a,Kliem2014a}. During catastrophe, magnetic free energy is always released by both magnetic reconnection and the work done by Lorentz force \citep{Chen2007a,Zhang2016a}. It was also demonstrated in previous studies that catastrophe has close relationship with instabilities \citep[e.g.][]{Demoulin2010a,Kliem2014a}.
\par
In previous studies, a 2.5 dimensional ideal MHD model in Cartesian coordinates was used to investigate the evolution of the equilibrium states associated with a flux rope embedded in bipolar magnetic fields. It was found that no catastrophe occurs for the flux rope of finite cross section in a completely closed bipolar configuration \citep{Hu2000a}, consistent with the conclusion in analytical analyses \citep{Forbes1991a,Forbes1995a}. If the background bipolar field is partially open, however, the magnetic system is catastrophic \citep{Hu2001a}. The equilibrium solutions are then bifurcated: the flux rope may either stick to the photosphere (lower branch solution) or be suspended in the corona  (upper branch solution). If the control parameter exceeds a critical value, the flux rope jumps upward from the lower branch to the upper branch, which is called ``upward catastrophe'' \citep{Zhang2016a}. Here control parameters characterize physical properties of the magnetic system; any parameter can be selected as the control parameter provided that different values of this parameter will result in different equilibrium states \citep{Kliem2014a,Zhang2016a}. Whether a system is catastrophic depends on how its equilibrium states evolve with the control parameter. Recently, \cite{Zhang2016a} found that there also exists a ``downward catastrophe'', i.e., a sudden jump from the upper branch to the lower branch, during which magnetic energy is also released, implying that the downward catastrophe might be a possible mechanism for energetic but non-eruptive activities, such as confined flares \citep[e.g.][]{Liu2014a,Yang2014a}, but observational evidence is being sought.
\par
Since catastrophe could account for many different solar activities, it is important to investigate what influences the existence and properties of the catastrophe. Previous studies have demonstrated that whether the background bipolar field is completely closed or partially open greatly influences the catastrophic behavior of the flux rope system. A question arises as to whether this is the only affecting factor . Due to the limit of current observing technologies, coronal magnetic configurations, corresponding to the background fields around the flux rope, can not be directly measured. What can be observed is the photospheric magnetic conditions. To reveal the influence of the photospheric magnetic conditions on the catastrophe of a flux rope system could not only help to better understand the decisive factors for catastrophe, but also shed light on the flare/CME productivity of active regions \citep[e.g.][]{Romano2007a,Schrijver2007a,Wang2008a,Chen2012a,Liu2016a}. By numerical simulations in spherical coordinates, \cite{Sun2007a} found that if the global photospheric flux is concentrated too close to the magnetic neutral line, the system losses its catastrophic behavior. Many solar eruptive activities originate from active regions \citep{Su2007a,Chen2011b,Sun2012a,Titov2012a}, the spatial scale of which is small as compared with the solar radius, hence Cartesian coordinates suit the simulations of flux ropes in active regions. In this paper, we use the same 2.5 dimensional ideal MHD model in Cartesian coordinates as in previous studies (\sect{sec:equation}) to simulate the evolution of the equilibrium states under different photospheric conditions with the background field either partially open (\sect{sec:presults}) or completely closed (\sect{sec:cresults}). Finally, a discussion about the implications of the simulation results is given in \sect{sec:discuss}.

\section{Basic equations and the initial and boundary conditions}
\label{sec:equation}
A Cartesian coordinate system is used and a magnetic flux function $\psi$ is introduced to denote the magnetic field as follows:
\begin{align}
\textbf{B}=\triangledown\times(\psi\hat{\textbf{\emph{z}}})+B_z\hat{\textbf{\emph{z}}}.\label{equ:mf}
\end{align}
Neglecting the radiation and heat conduction in the energy equation, the 2.5-D MHD equations can be written in the non-dimensional form:
\begin{align}
&\frac{\partial\rho}{\partial t}+\triangledown\cdot(\rho\textbf{\emph{v}})=0,\label{equ:cal-st}\\
&\frac{\partial\textbf{\emph{v}}}{\partial t}+\textbf{\emph{v}}\cdot\triangledown\textbf{\emph{v}}+\triangledown T +\frac{T}{\rho}\triangledown\rho+\frac{2}{\rho\beta_0}(\vartriangle\psi\triangledown\psi+B_z\triangledown B_z+\triangledown\psi\times\triangledown B_z)+g\hat{\textbf{\emph{y}}}=0,\\
&\frac{\partial\psi}{\partial t}+\textbf{\emph{v}}\cdot\triangledown\psi=0,\\
&\frac{\partial B_z}{\partial t}+\triangledown\cdot(B_z\textbf{\emph{v}})+(\triangledown\psi\times\triangledown v_z)\cdot\hat{\textbf{\emph{z}}}=0,\\
&\frac{\partial T}{\partial t}+\textbf{\emph{v}}\cdot\triangledown T +(\gamma-1)T\triangledown\cdot\textbf{\emph{v}}=0,\label{equ:cal-en}
\end{align}
where $\rho, \textbf{\emph{v}}, T, \psi$ denote the density, velocity, temperature and magnetic flux function, respectively; $B_z$ and $v_z$ correspond to the z-component of the magnetic field and the velocity, which are parallel to the axis of the flux rope; $g$ is the normalized gravity, $\beta_0=2\mu_0\rho_0RT_0L_0^2/\psi_0^2=0.1$ is the characteristic ratio of the gas pressure to the magnetic pressure, where $\mu_0$ and $R$ is the vacuum magnetic permeability and gas constant, respectively; $\rho_0=3.34\times10^{-13}\mathrm{~kg~m^{-3}}$, $T_0=10^6\mathrm{~K}$, $L_0=10^7\mathrm{~m}$, and $\psi_0=3.73\times10^3\mathrm{~Wb~m^{-1}}$ are the characteristic values of density, temperature, length and magnetic flux function, respectively. The initial corona is isothermal and static with
\begin{align}
T_c\equiv T(0,x,y)=1\times10^6 ~\mathrm{K},\ \  \rho_c\equiv\rho(0,x,y)=\rho_0\mathrm{e}^{-gy}.
\end{align}
\par
In this paper, the background field is taken to be bipolar, either partially open or completely closed (see Sections \ref{sec:presults} and \ref{sec:cresults} for details). It is assumed to be symmetrical relative to the $y$-axis. The lower boundary $y=0$ corresponds to the photosphere; $\psi$ at the lower boundary is always fixed at the value of the background field except during the emergence of the flux rope. There is a positive and a negative surface magnetic charge located at the photosphere within $-b<x<-a$ and $a<x<b$, respectively. The photospheric magnetic flux distribution is characterized by the distance $d$ between the inner edges of the two charges ($d=2a$) and the width $w$ of the charges ($w=b-a$). With different values of $d$ and $w$, different background configurations can be calculated by complex variable methods accordingly (see Sections \ref{sec:presults} and \ref{sec:cresults}).
\par
The magnetic properties of the flux rope are characterized by the axial magnetic flux passing through the cross section of the flux rope, $\Phi_z$, and the annular magnetic flux of the rope of per unit length along $z$-direction, $\Phi_p$, which is simply the difference in $\psi$ between the axis and the outer boundary of the flux rope. Here we select $\Phi_z$ as the control parameter, i.e. we analyze the evolution of the equilibrium solutions of the system versus $\Phi_z$ with a fixed $\Phi_p$. The varying $\Phi_z$ represents an evolutionary scenario, e.g., flux emergence \citep{Archontis2008a} or flux-feeding from chromospheric fibrils \citep{Zhang2014a}. It should be noted that, if not changed manually, $\Phi_z$ and $\Phi_p$ of the rope should be maintained to be conserved, which is achieved by the numerical techniques proposed by \cite{Hu2003a}.
\par
With the initial conditions, equations (\ref{equ:cal-st}) to (\ref{equ:cal-en}) are solved by the multi-step implicit scheme \citep{Hu1989a} to allow the system to evolve to equilibrium states. In order to investigate the influence of the photospheric magnetic conditions on the catastrophic behavior of the flux rope system, we calculate the evolution of the flux rope in different background configurations in the following procedures. Starting from a background configuration with given $d$ and $w$, we let a magnetic flux rope emerge from the central area of the base. 
Following \cite{Hu2000a} and \cite{Hu2001a}, the emergence of the flux rope is assumed to begin at $t=0$ in the central area of the base and end at $t=\tau_E=87$ s, after which the flux rope are fully detached from the base. The emerging speed is uniform, and then the emerged part of the flux rope is bounded by $x=\pm x_E$ at time $t$, where
\begin{align}
x_E=(a^2-h_E^2)^{1/2},~h_E=a(2t/\tau_E-1),
\end{align}
and $a=5$ Mm is the radius of the rope. The relevant parameters at the base of the emerged part of the flux rope ($y=0, |x|\leq x_E$) are specified as:
\begin{align}
&\psi(t,x,0)=\psi_0(x,0)+\psi_E(t,x),\\
&\psi_E(t,x)=\frac{C_E}{2}\mathrm{ln}\left(\frac{2a^2}{a^2+x^2+h_E^2}\right),\\
&B_z(t,x,0)=C_Ea(a^2+x^2+h_E^2)^{-1},\\
&v_y(t,x,0)=2a/\tau_E,~v_x(t,x,0)=v_z(t,x,0)=0,\\
&T(t,x,0)=2\times10^4\mathrm{~K},~\rho(t,x,0)=1.67\times10^{-11}\mathrm{~kg~m^{-3}},
\end{align}
where $\psi_0$ is the flux function of the background field, and $C_E$ is a constant controlling the initial magnetic properties of the emerged rope. The values of $C_E$ range from 2.0 to 4.0 for different cases. The outer boundary of the emerged rope is determined by $\psi=\psi_{x=0,y=0}.$ After the emergence of the rope, we obtain an equilibrium state with the flux rope sticking to the lower boundary. Starting from such a state, new equilibrium solutions with different $\Phi_z$ but the same $\Phi_p=\Phi_p^0$ are calculated, and thus we obtain the evolution of the flux rope in equilibrium states as a function of $\Phi_z$ in the given background configuration, as  described by the geometric parameters of the flux rope, including the height of the rope axis, $H$, and the length of the current sheet below the rope, $L_c$. Similar procedures are repeated for background configurations with different $d$ and $w$ to obtain the evolutionary profiles of the flux rope under different photospheric flux distributions. The influence of the photospheric conditions could then be revealed by comparing the evolutions of the flux rope under different background configurations (see \sect{sec:presults} and \sect{sec:cresults}). Note that since we adjust $\Phi_z$ in our simulation to calculate different equilibrium solutions, the value of $C_E$ is insignificant, which only influences the initial magnetic properties.
\par
If the flux rope breaks away from the photosphere, a vertical current sheet will form beneath it. In our numerical scheme, any reconnection will reduce the value of $\psi$ at the reconnection site. Therefore, by keeping $\psi$ invariant along the newly formed current sheet, reconnections, including both numerical and physical magnetic reconnections, are completely prevented across the current sheet.
\par

\section{Simulation results in partially open bipolar field}
\label{sec:presults}
First we analyze the influence of photospheric flux distributions on the magnetic system associated with partially open bipolar background fields. Following \cite{Hu2001a}, the background magnetic field can be cast in the complex variable form
\begin{align}
f(\omega)\equiv B_x-iB_y=\frac{(\omega+iy_N)^{1/2}(\omega-iy_N)^{1/2}}{F(a,b,y_N)}\mathrm{ln}\left( \frac{\omega^2-a^2}{\omega^2-b^2}\right),
\end{align}
where $\omega=x+iy$, the position of the neutral point of the partially open bipolar field is ($y=y_N$, $x=0$), and
\begin{align}
\nonumber &F(a,b,y_N)=\frac{1}{b-a}\int_a^b(x^2+y_N^2)^{1/2}dx=\frac{1}{2(b-a)}\times\\ &\left[b(b^2+y_N^2)^{1/2}-a(a^2+y_N^2)^{1/2}+y_N^2\mathrm{ln}\left(\frac{b+(b^2+y_N^2)^{1/2}}{a+(a^2+y_N^2)^{1/2}} \right)\right].
\end{align}
The magnetic flux function is then calculated by
\begin{align}
\psi(x,y)=\mathrm{Im}\left\lbrace\int f(\omega)d\omega \right\rbrace,\label{equ:complex}
\end{align}
and the flux function at the photosphere can be derived as
\begin{equation}
 \psi(x,0) = \left\{
              \begin{array}{ll}
              {\psi_c}, &{|x|<a}\\
              {\psi_c F(|x|,b,y_N)/F(a,b,y_N)}, &{a\leqslant|x|\leqslant b}\\
              {0}, &{|x|>b}
              \end{array}  
         \right.
\end{equation}
where $\psi_c=(b-a)\pi=\pi w$ is the total magnetic flux emanating upward from the positive charge per unit length along the z-axis. Note that $\psi_c$ is independent of the distance $d$.
\par
\begin{figure*}
\includegraphics[width=\hsize]{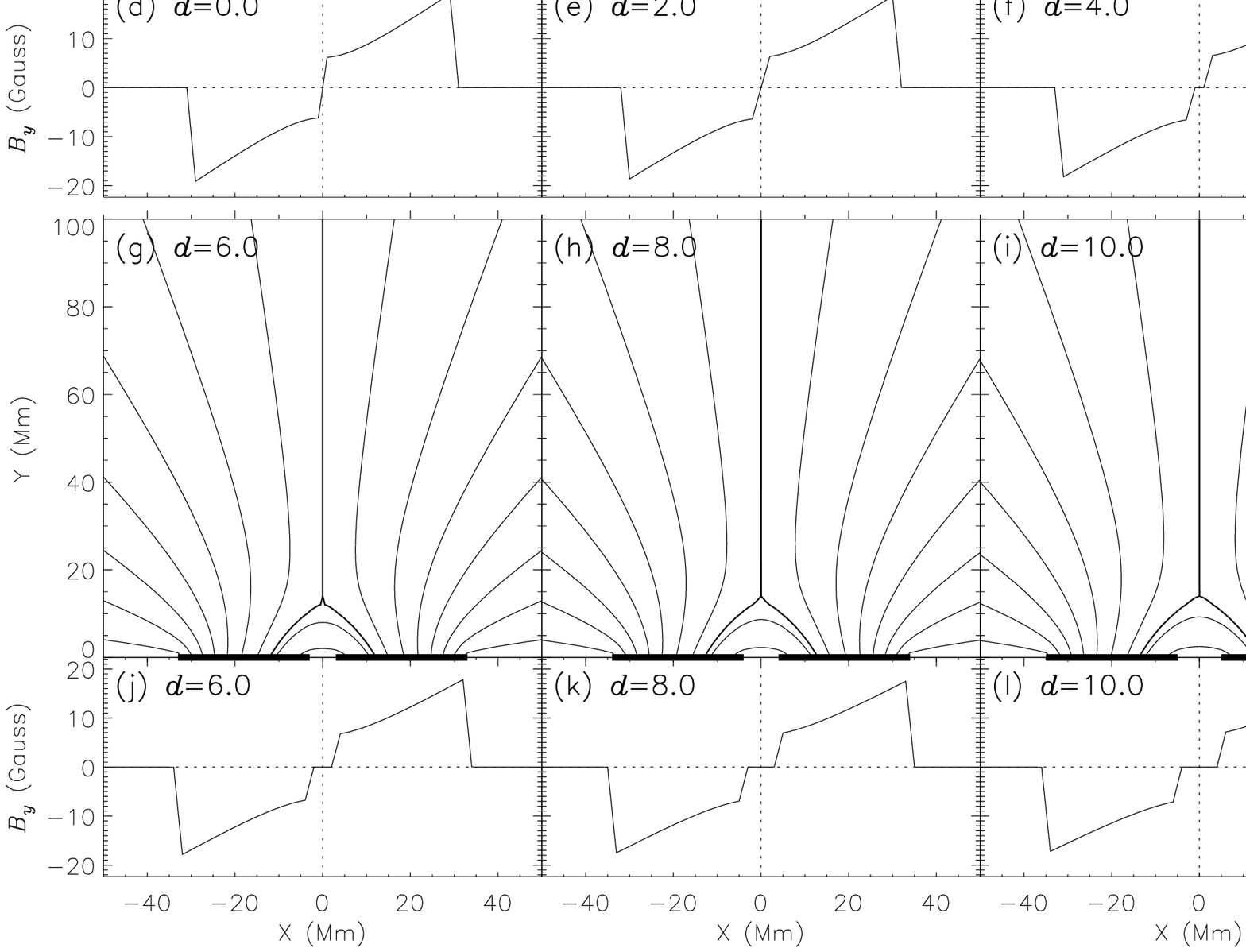}
\caption{The partially open bipolar background configurations and the corresponding radial components of the magnetic field ($B_y$) at the photosphere ($y=0$) for different $d$, which is selected to be $0.0, 2.0, 4.0, 6.0, 8.0, 10.0$ Mm, respectively; $w$ is 30 Mm for all the six cases. The two surface magnetic charges for different cases are marked by the black solid lines at $y=0$ in panels (a)-(c) and (g)-(i).}\label{fig:inits}
\end{figure*}
\begin{figure*}
\includegraphics[width=\hsize]{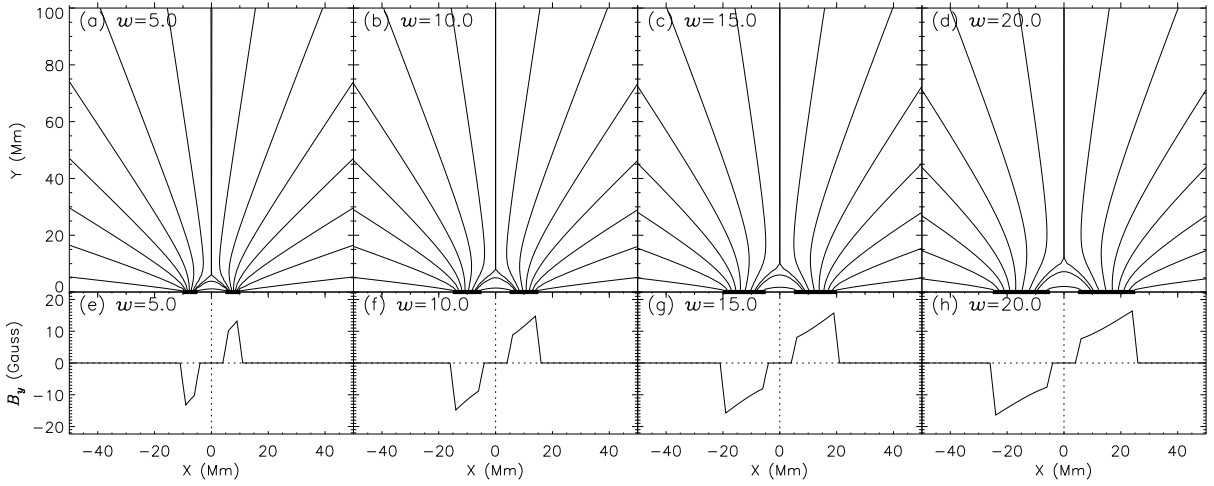}
\caption{The partially open bipolar background configurations and the corresponding radial components of the magnetic field ($B_y$) at the photosphere ($y=0$) for different $w$, which is selected to be $5.0, 10.0, 15.0, 20.0$ Mm, respectively; $d$ is 10 Mm for all the four cases. The two surface magnetic charges for different cases are marked by the black solid lines at $y=0$ in panels (a)-(d).}\label{fig:initc}
\end{figure*}

The magnetic configurations of the background fields are shown in \fig{fig:inits} and \fig{fig:initc}. The two magnetic surface charges are denoted by the thick lines in the figures. \fig{fig:inits}(a)-\ref{fig:inits}(c) and \ref{fig:inits}(g)-\ref{fig:inits}(i) show the background field configurations for $d=0.0$, $2.0$, $4.0$, $6.0$, $8.0$, $10.0$ Mm, respectively, with the same $w=30$ Mm, whereas \fig{fig:initc}(a)-\ref{fig:initc}(d) for $w=5.0$, $10.0$, $15.0$, $20.0$ Mm with the same $d=10.0$ Mm. The corresponding photospheric distributions of the normal component of the magnetic field, $B_y$, are plotted in \fig{fig:inits}(d)-\ref{fig:inits}(f), \ref{fig:inits}(j)-\ref{fig:inits}(l), and \fig{fig:initc}(e)-\ref{fig:initc}(h), respectively. The ratio of the magnetic flux of the open component to the total flux of the background field is determined by
\begin{align}
\alpha=\frac{\psi_N}{\psi_c},
\end{align}
where 
\begin{align}
\psi_N=\frac{\pi(b^2-a^2)}{2F(a,b,y_N)}
\end{align}
is the flux function at the neutral point $y=y_N$, corresponding to the flux of the open component. For the background fields with different $d$ and $w$, $\alpha$ is always selected to be 0.8, and the resultant $y_N$ varies slightly among different cases. The computational domain is taken to be $0<x<100$ Mm, $0<y<300$ Mm, with symmetrical condition used for the left side ($x=0$). As mentioned above, $\psi$ at the lower boundary is always fixed to be $\psi_0$ except during the emergence of the flux rope. In the simulation, potential field conditions are used at the top ($y=300$ Mm) and right ($x=300$ Mm) boundaries, except for the location of the current sheet ($x=0$ Mm, $y=300$ Mm), at which increment-equivalue extrapolation is used.
\par
\begin{figure*}
\includegraphics[width=\hsize]{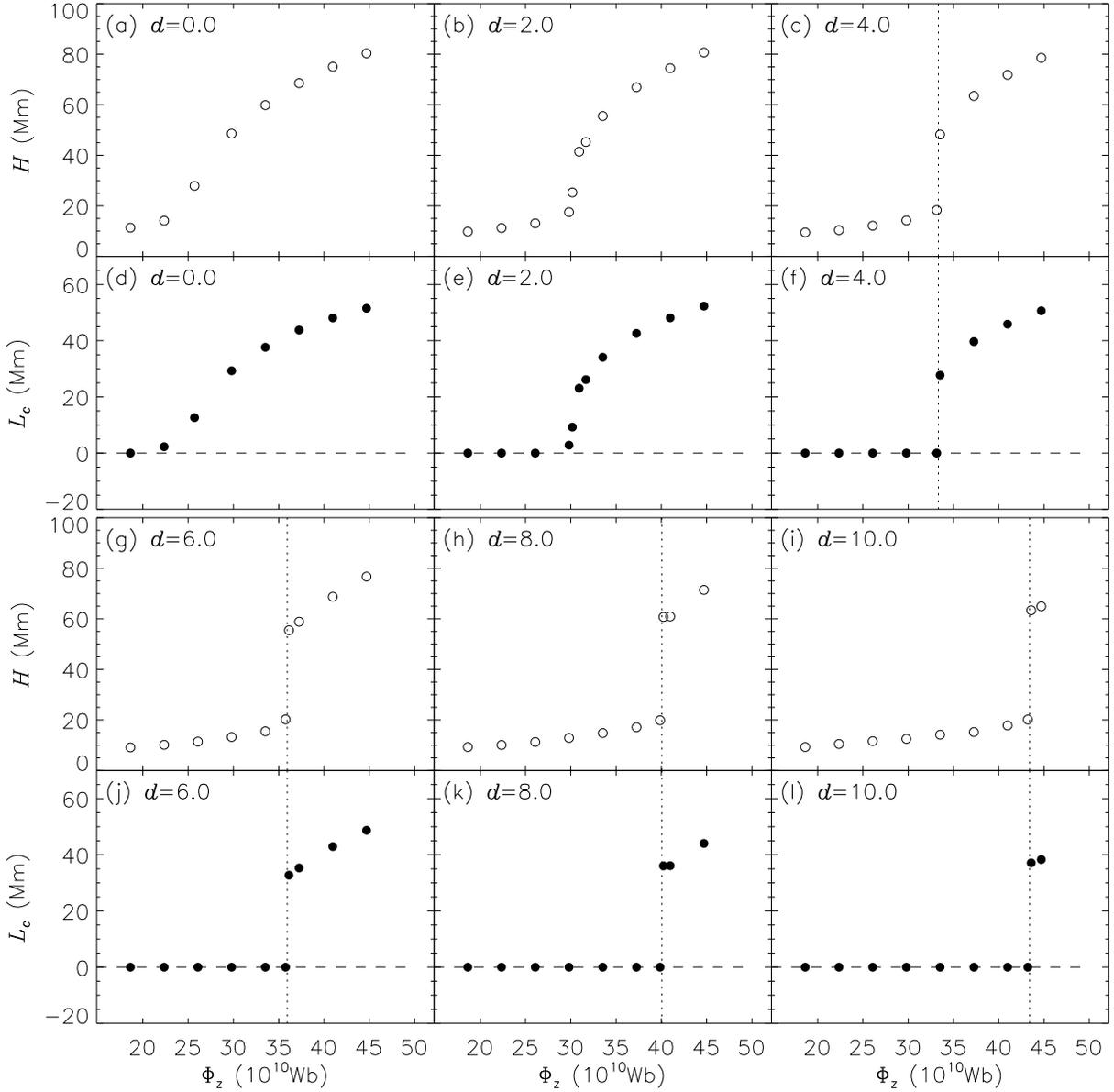}
\caption{The height of the flux rope axis ($H$) and the length of the current sheet below the rope ($L_c$) are shown as functions of the control parameter $\Phi_z$ for partially open bipolar background fields with different $d$; $\Phi_p$ is selected to be 1.49$\times10^4$ Wb m$^{-1}$ for all the equilibrium solutions. The evolutions of $H$ are plotted by small black circles, and those of $L_c$ by black dots. The vertical dotted lines represent the catastrophic points of the catastrophic cases.}\label{fig:fluxs}
\end{figure*}
\begin{figure*}
\includegraphics[width=\hsize]{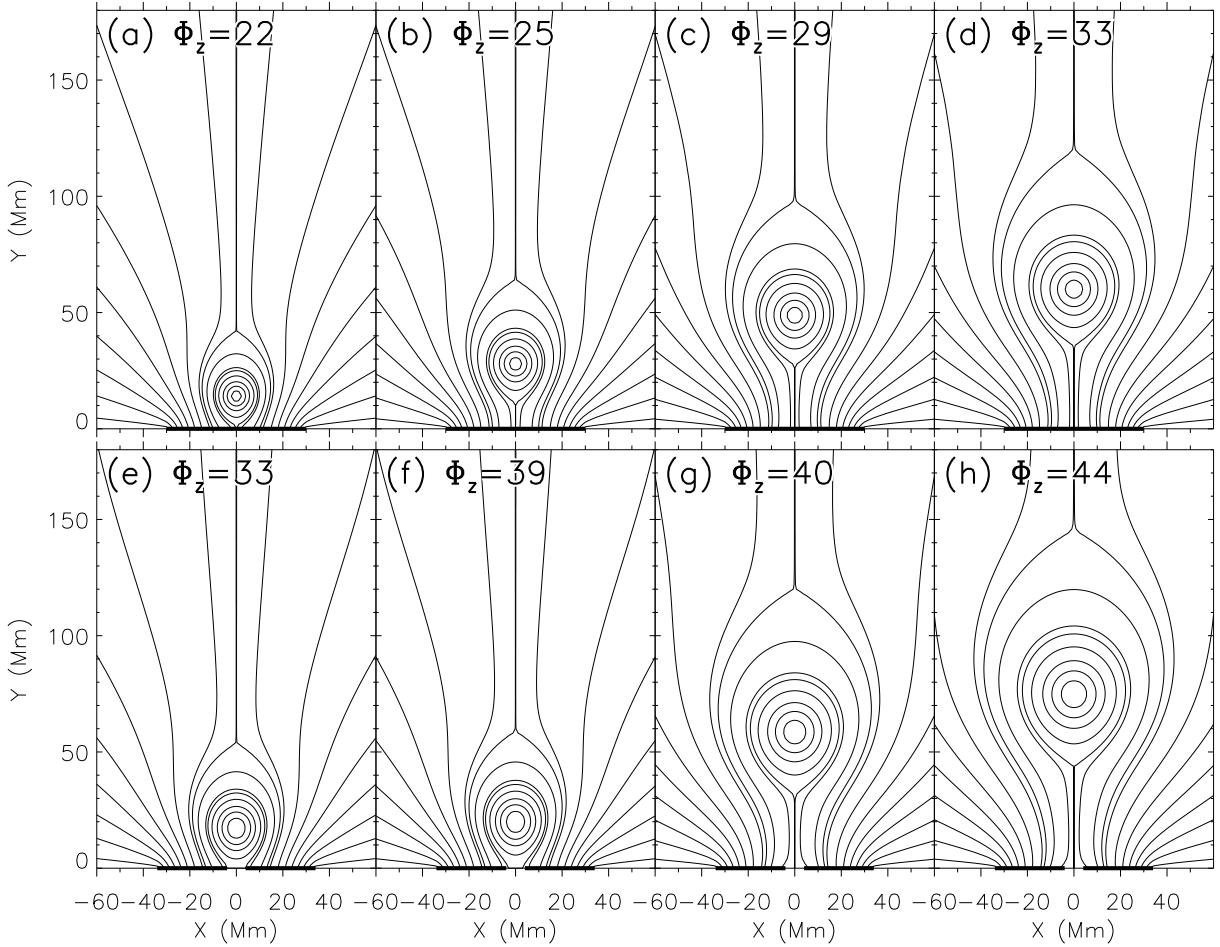}
\caption{Evolution of a non-catastrophic magnetic system ($d=0.0$ Mm) in top panels and that of a catastrophic case ($d=8.0$ Mm) in bottom panels. $\Phi_z$ is in units of $10^{10}$ Wb.}\label{fig:evo}
\end{figure*}

By the simulating procedures introduced in \sect{sec:equation}, the evolutions of the equilibrium states of the system consisting of a flux rope in the background configurations with different $d$ are calculated, as plotted in \fig{fig:fluxs}. \fig{fig:fluxs}(a)-\ref{fig:fluxs}(c) and \ref{fig:fluxs}(g)-\ref{fig:fluxs}(i) show the evolutions of $H$ as a function of $\Phi_z$, and \fig{fig:fluxs}(d)-\ref{fig:fluxs}(f) and \ref{fig:fluxs}(j)-\ref{fig:fluxs}(l) show those of $L_c$. The equilibrium solutions with different values of the control parameter $\Phi_z$ are represented by the circles (for $H$) and dots (for $L_c$). $\Phi_p$ of the flux rope for all equilibrium solutions in \fig{fig:fluxs} is 1.49$\times10^4$ Wb m$^{-1}$. For all of the 6 cases, the flux rope sticks to the photosphere at first ($\Phi_z=18.6\times10^{10}$ Wb); as $\Phi_z$ increases, the flux rope breaks away from the base and levitates in the corona. The transitions between these two different kinds of equilibrium states, however, are quite different for different values of $d$. For $d=0.0$ Mm, corresponding to panels (a) and (d) in \fig{fig:fluxs}, both $H$ and $L_c$ increase continuously with increasing $\Phi_z$; no catastrophe takes place. The magnetic configurations of the equilibrium states with different $\Phi_z$ in this case are plotted in the top panels in \fig{fig:evo}. For $d=2.0$ Mm, although the variations of $H$ and $L_c$ versus $\Phi_z$ are steeper, the transition from sticking to the photosphere ($L_c=0$) to levitating in the corona ($L_c>0$) is still continuous, indicating that no catastrophe takes place either. For $d\geqslant4.0$ Mm, however, the equilibrium states are diverged into two branches and the flux rope suddenly jump upward as soon as $\Phi_z$ reaches a critical value, resulting in a discontinuous transition between the two branches of equilibrium states. Catastrophe takes place under these background configurations, and the critical value $\Phi_z^c$ at which catastrophe takes place is called catastrophic point, marked by the vertical dotted lines in \fig{fig:fluxs}. The magnetic configurations of the equilibrium states of the magnetic system with $d=8.0$ Mm are plotted in the bottom panels in \fig{fig:evo}. As shown in the figure, the flux rope keeps sticking to the photosphere before reaching the catastrophic point $\Phi_z^c=40\times10^{10}$ Wb, and then jumps upward and levitates in the corona after reaching $\Phi_z^c$. Note that steep transition is different from catastrophe in essence. Steep transition is still continuous, so that variations of the control parameter resulting from disturbances could only trigger movements of the flux rope in a spatial scale comparable to the disturbance itself, no matter how steep the transition is. In contrast, catastrophe manifests as a discontinuous jump, so that even an infinitesimal enhancement of the control parameter to reach the catastrophic point could trigger a catastrophe of the system, during which the flux rope jumps from the lower branch to the upper branch. Therefore, the spatial range of the resultant jump of catastrophe could be much larger than that of the disturbances. As shown in observations, the spatial range of eruptive activities, such as flares and CMEs, is much larger than that of photospheric or coronal disturbances \citep[e.g.][]{Priest1982book}, which are regarded as possible triggers for these eruptions. The tremendous difference in the spatial scales determines that only via catastrophe could small-scale disturbances trigger large-scale eruptive activities. Our simulations reveal that, if the photospheric flux is concentrated too much toward the polarity inversion line (PIL) in the central area of the active region (i.e. $d$ is small enough), the system with a flux rope embedded in a partially open bipolar field possesses no catastrophe.
\par
The value of $d$ not only determines the existence of the catastrophe, but also influences the properties of the catastrophe. The parameters of the catastrophes under background configurations with different $d$ are tabulated in \tbl{tbl:s}. For larger $d$, the catastrophic point $\Phi_z^c$ is higher. This might result from the stronger constraint exerted by the background field with larger $d$. The spatial amplitude of the catastrophe also increases with $d$. Moreover, we calculate the magnetic energy per unit length in $z$ direction within the domain by
\begin{align}
E=\int\int\frac{B^2}{2\mu_0}dxdy.
\end{align}
Following \cite{Zhang2016a}, the variation of $E$ could shed light on the evolution of the magnetic energy of the whole magnetic system semi-quantitatively. As shown in \tbl{tbl:s}, more magnetic energy is released in the case with larger $d$. Since there is no magnetic reconnection in our simulation, magnetic energy should mainly be released via the work done by the Lorentz force \citep{Zhang2016a}, which is also called Amp\`{e}re's force in some papers. For catastrophic case under larger $d$, the higher $\Phi_z^c$ corresponds to stronger magnetic field in the flux rope when the catastrophe takes place, so that the Lorentz force dominating the catastrophe is also stronger. Moreover, the larger amplitude of the catastrophe under larger $d$ indicates more drastic evolution of the system. Therefore, the work done by Lorentz force should be larger, so that more magnetic energy is released. 
\par
\begin{figure*}
\includegraphics[width=\hsize]{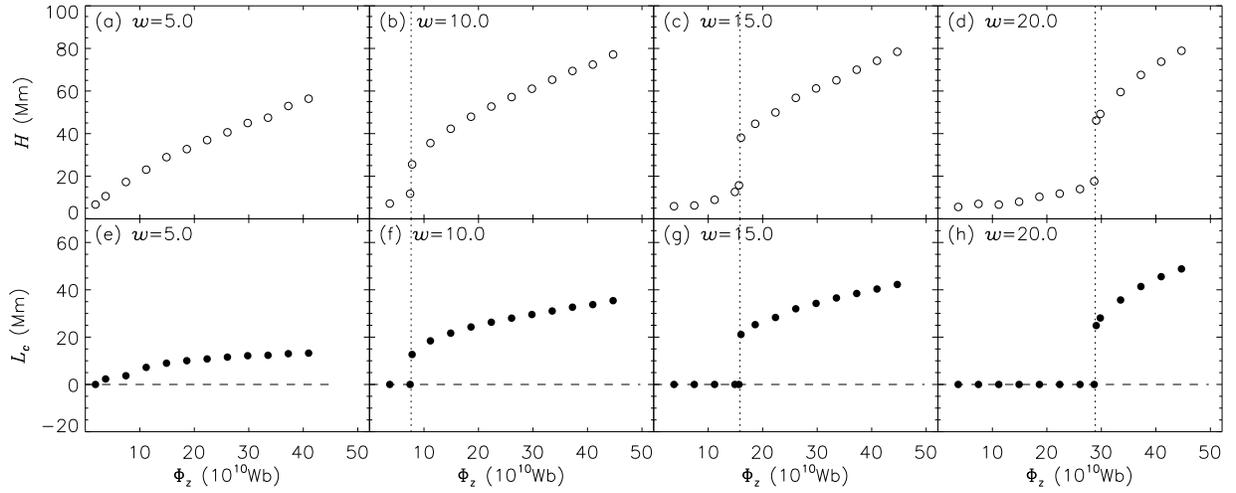}
\caption{$H$ and $L_c$ versus $\Phi_z$ for partially open bipolar background fields with different $w$. $\Phi_p$ is fixed at 2.24$\times10^3$ Wb m$^{-1}$ for $w=5.0$ Mm, whereas $\Phi_p=7.45\times10^3$ Wb m$^{-1}$ for the ohter cases. The meanings of the symbols are the same as those in \fig{fig:fluxs}. }\label{fig:fluxc}
\end{figure*}

The evolutions of the equilibrium solutions of the system under different $w$ are plotted in \fig{fig:fluxc}. The meanings of the symbols in \fig{fig:fluxc} are the same as those in \fig{fig:fluxs}. $\Phi_p$ of the flux rope is fixed to be 2.24$\times10^3$ Wb m$^{-1}$ for the case with $w=5.0$ Mm, and 7.45$\times10^3$ Wb m$^{-1}$ for $w=10.0, 15.0, 20.0$ Mm. Similarly, the transition from the state with the flux rope sticking to the photosphere to that with the rope levitating in the corona varies with $w$. For the case in which the flux rope is embedded in the background field with $w=5.0$ Mm, the flux rope in equilibrium state evolves continuously from sticking to the photosphere to levitating in the corona with increasing $\Phi_z$, indicating that there is no catastrophe. For the cases with $w\geqslant10.0$ Mm, the equilibrium solutions are separated into two branches and the catastrophe takes place under these configurations, namely, the flux rope suddenly jumps upward at the catastrophic point, manifested as a discontinuous transition from the lower branch to the upper branch. Thus we conclude that small enough $w$ of the background field might also result in a non-catastrophic system. For catastrophic cases, different values of $w$ also influence the properties of the catastrophe, which are tabulated in \tbl{tbl:c}. Similarly, larger $w$ of the background field results in higher catastrophic point, larger amplitude of the catastrophe, and more released magnetic energy. The influence of $w$ on photospheric magnetic conditions is complex. Since the total flux $\psi_c=\pi w$, a smaller $w$ corresponds to a less total magnetic flux, resulting in a weaker background field. This indicates that the non-catastrophic case with small enough $w$ also has very weak photospheric regions of the background field. On the other hand, a smaller $w$ also results in a smaller distance between the weighted centers of the two surface charges, which is similar as the influence of decreasing $d$. It should be noted that \cite{Forbes1995a} found that a magnetic system with point photospheric sources (i.e. $w=0$) is catastrophic. This discrepancy results from the differences in the models used in \cite{Forbes1995a} and our simulation: in our model, if $w$ approaches 0, $\psi_c$ also vanishes, whereas $\psi_c$ is finite in \cite{Forbes1995a} with $w=0$. 
\par
In summary, catastrophe does not always exist in the magnetic system consisting of a flux rope embedded in a partially open bipolar field. Both the existence and the properties of the catastrophe are greatly influenced by the photospheric magnetic flux distribution of the background field.

\section{Simulation results in completely closed background field}
\label{sec:cresults}
\begin{figure*}
\includegraphics[width=\hsize]{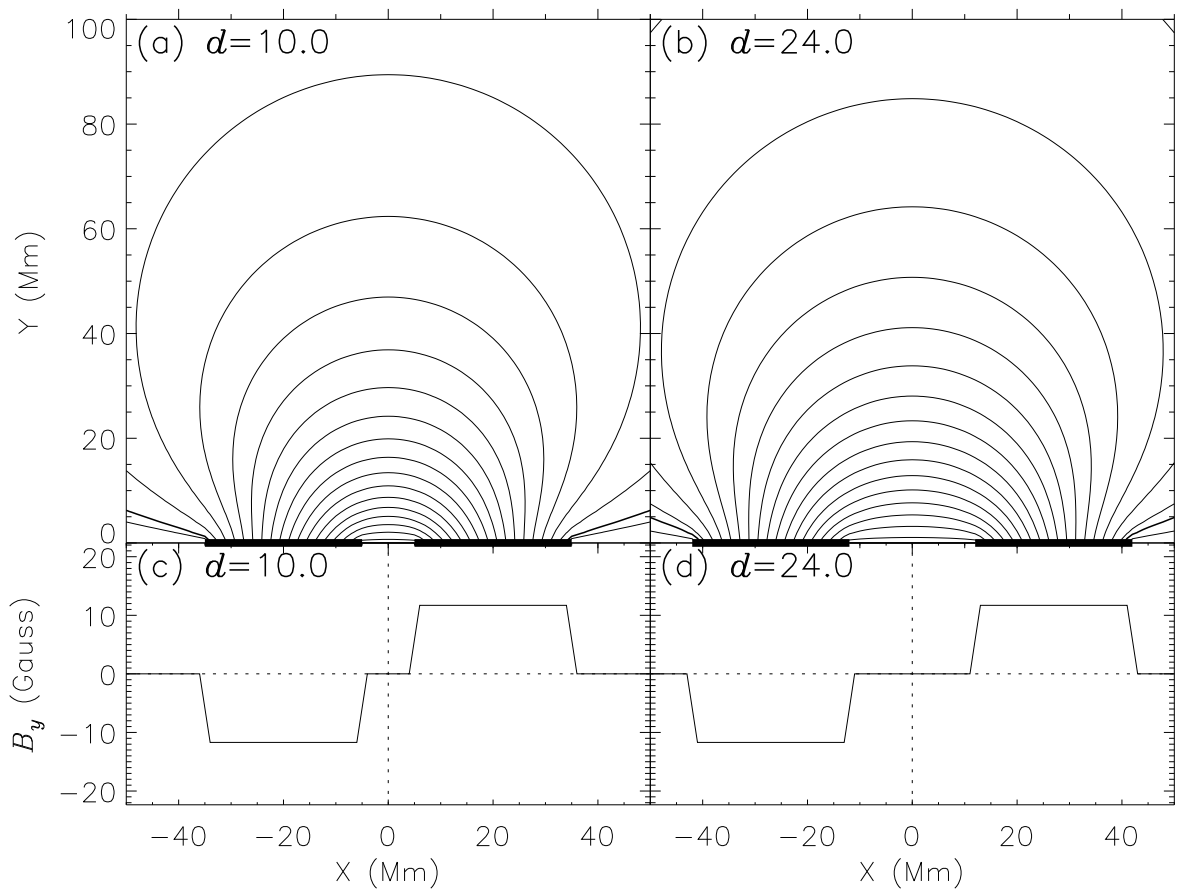}
\caption{The completely closed bipolar background configurations and the corresponding radial components of the magnetic field ($B_y$) at the photosphere ($y=0$) for $d=10$ Mm in left panels and $d=24$ Mm in right panels. $w$ is 30 Mm for both the two cases. The two surface magnetic charges for different cases are marked by the black solid lines at $y=0$ in panels (a) and (b).}\label{fig:initclo}
\end{figure*}

For completely closed bipolar background field, we calculated two typical cases to investigate the influence of photospheric conditions. Following \cite{Hu2000a}, the potential background field can be cast in
\begin{align}
f(\omega)\equiv B_x-iB_y=\mathrm{ln}\left( \frac{\omega^2-a^2}{\omega^2-b^2}\right),
\end{align}
and then the flux function is also calculated by equation (\ref{equ:complex}). The background configurations and the corresponding $B_y$ at the photosphere are shown in \fig{fig:initclo}, where $d=10.0$ Mm for left panels and 24.0 Mm for right panels, respectively; $w$ is fixed to be 30 Mm for both cases. The initial and boundary conditions for completely closed background field slightly differ from those for partially open ones. Improper boundary conditions might open the closed arcade near the top of the computational domain during the simulation, which will result in a partially open background configuration. In order to investigate the characteristics of catastrophe in completely closed bipolar field, the background configuration must be guaranteed to be always purely closed during the whole simulation. To achieve this, the top and right boundaries are fixed during the simulation. Following \cite{Zhang2016a}, we enlarge the computational domain to $0<x<200$ Mm, $0<y<300$ Mm, so as to minimize the influence of the boundary conditions. Moreover, for stability and simplicity of the simulation, a relaxation method is used to obtain force-free equilibrium solutions, which involves resetting the temperature and density in the computational domain to their initial values, so that the pressure gradient force is always balanced everywhere by the gravitational force \citep{Hu2004a}.
\par
\begin{figure*}
\includegraphics[width=\hsize]{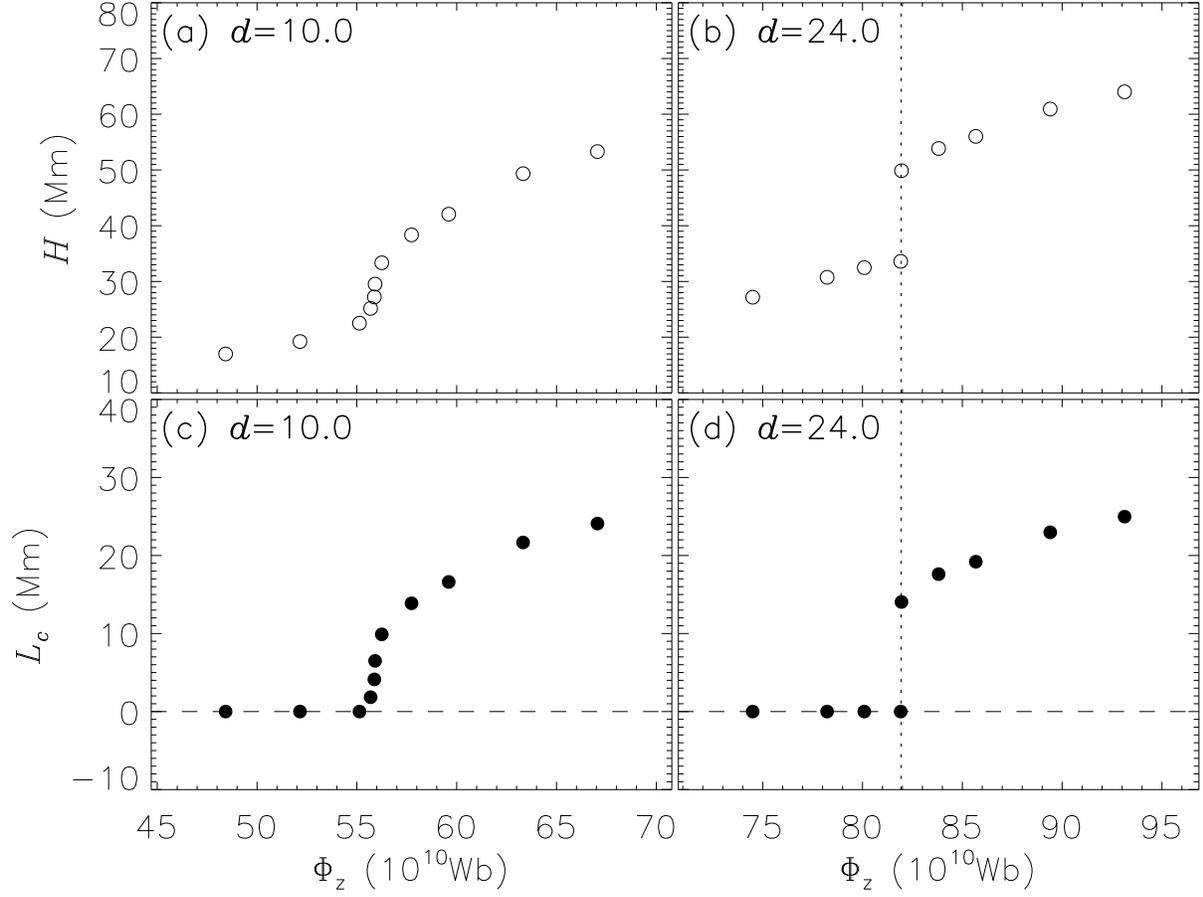}
\caption{$H$ and $L_c$ versus $\Phi_z$ for completely closed bipolar background fields with different $d$; $\Phi_p$ is selected to be 2.98$\times10^4$ Wb m$^{-1}$ for all the equilibrium solutions. The meanings of the symbols are the same as those in \fig{fig:fluxs}. }\label{fig:fluxclo}
\end{figure*}

By the simulating procedures introduced in \sect{sec:equation}, the evolutions of the equilibrium solutions under different background fields are simulated, as shown in \fig{fig:fluxclo}. $\Phi_p$ for all the equilibrium states in \fig{fig:fluxclo} is selected to be 2.98$\times10^4$ Wb m$^{-1}$. For the first case with $d=10.0$ Mm, both $H$ and $L_c$ increase continuously and monotonously with increasing $\Phi_z$; no catastrophe takes place. This result is consistent with \cite{Hu2000a}, in which the evolution of the magnetostatic equilibrium solutions under the same photospheric condition is simulated. The evolution of the system with $d=24.0$ Mm, however, shows an obvious catastrophic behavior: the flux rope keeps sticking to the photosphere till $\Phi_z=8.20\times10^{11}$ Wb, at which the flux rope jumps upward and levitates in the corona, resulting in a discontinuous transition from the lower branch to the upper branch. From the simulation results, we conclude that the magnetic system consisting of a flux rope embedded in a completely closed bipolar field is not always non-catastrophic; under certain photospheric flux distributions, catastrophe could take place with increasing control parameters. The influence of the photospheric condition on the catastrophe of the system in completely closed bipolar configuration is similar as that on the system in partially open configuration: large $d$ favours the existence of catastrophe.

\section{Discussion and Conclusion}
\label{sec:discuss}
To investigate the influence of the photospheric magnetic conditions on the catastrophe of the flux rope system in active regions, we simulate the evolution of the equilibrium states associated with a flux rope in a partially open or a completely closed bipolar background fields with different photospheric magnetic conditions. For the partially open bipolar configuration, it is found that both the distance $d$ between the two magnetic surface charges located at the photosphere and their width $w$ influence the catastrophe of the rope system. The catastrophe could only take place when $d$ and $w$ of the background field is not very small, namely, the photospheric flux is not concentrated too much toward the central area and the source regions of the bipolar field are not too weak. If either $d$ or $w$ is small enough, the flux rope evolves continuously with increasing $\Phi_z$, i.e., there is no catastrophe under this configuration. Moreover, photospheric magnetic conditions also affect the properties of the catastrophe. The larger $d$ of the background field, the higher the catastrophic point, the larger the amplitude the catastrophe, and the more magnetic energy is released during the catastrophe. The catastrophic evolution of the system is more intense under larger value of $d$. Similar conclusions hold for $w$. For completely closed bipolar configuration, it is also found that there is no catastrophe in the magnetic system under the photospheric condition with small $d$, whereas catastrophe takes place for large $d$. 
\par
It is demonstrated that the evolution of the flux rope system is strongly influenced by photospheric magnetic conditions. As mentioned above, only the magnetic conditions at the photosphere can be directly obtained in observations. Our simulation results may have significant implications for the relationship between the properties of active regions and the productivity of flares and CMEs, as well as the intensity of these eruptive cases. Long-term evolution of active regions can be divided into six evolutionary phases \citep{Tapping2001a,vanDriel2015a}: (1) $Emergence$, (2) $Growth$, (3) $Maximum~development$, (4) $Early~decay$, (5) $Late~decay$, and (6) $Remnant$. At the $Emergence$ phase, active regions usually appear as small, compact, bipolar plages (small $w$). From our simulation results, we may infer that the magnetic systems in active regions trend to be non-catastrophic at the $Emergence$ phase. At the $Growth$ phase, flux emergence proceeds vigorously, so that $w$ increases, which might correspond to the catastrophic cases in our simulations. By using the full-disk magnetograms and H$\alpha$ observations over the period 1\verb"-"13 November 1981, \cite{Tapping2001a} analyzed the evolutions of several active regions, and concluded that the flare index, a parameter describing the flare productivity of an active region, peaks strongly at the $Growth$ phase, consistent with the prediction from our simulation results. Thus we suggest that the peak of the flare index at the $Growth$ phase might result from the influence of the photospheric flux distributions on the catastrophe of the magnetic systems in active regions.
\par
Our simulation results reveal that whether the background bipolar field is completely closed or partially open is not the only determinant of the existence of catastrophe. Under certain photospheric conditions, catastrophe could not only take place in completely closed configuration but also cease to exist in partially open configuration. The openness of the bipolar field or the photospheric magnetic conditions actually result in different background configurations. Thus we may conclude that it is the configuration of the background field that determines whether catastrophe exists and influences the properties of the catastrophe of the system (if it exists); if different values of some parameter could result in different background configurations, this parameter might also affect the catastrophic behavior of the system.
\par
It should be noted that our approach is different from previous studies on catastrophe triggered by photospheric motions, such as \cite{Forbes1995a}, \cite{Hu2001b}. In those studies, the distance $d$ or the width $w$ is selected as the control parameter so that the changing control parameter represents the photospheric motions. It was found that either convergence (decreasing $d$) or shrinkage (decreasing $w$) of the photospheric source regions could trigger an upward catastrophe of the given system. In this paper, however, $d$ and $w$ are not control parameters; they characterize the photospheric flux distribution. Different values of $d$ and $w$ are selected to obtain different background fields, therefore representing different magnetic systems. For each system, we adjust $\Phi_z$, a property of the flux rope itself, to analyze whether the magnetic system is catastrophic. In essence, previous studies concern whether the given magnetic system is catastrophic under certain photospheric motions, whereas the present study intends to answer under what photospheric flux distributions the magnetic system is catastrophic with variations of the flux rope itself.
\par
This research is supported by Grants from NSFC 41131065, 41574165, 41421063, 41474151 and 41222031, MOEC 20113402110001, CAS Key Research Program KZZD-EW-01-4, and the fundamental research funds for the central universities WK2080000077. R.L. acknowledges the support from the Thousand Young Talents Program of China.


\begin{deluxetable}{cccccc}
\tablewidth{0pt}
\tablecaption{Parameters of the catastrophes under different $d$ with $w=30.0$ Mm\label{tbl:s}}
\tablehead{ \colhead{$d$(Mm)} & \colhead{$\Phi_z^c$($10^{10}$ Wb)} & \colhead{$\bigtriangleup L_c$(Mm)} & \colhead{$\bigtriangleup E$(J m$^{-1}$)} & \colhead{$E$(J m$^{-1}$)} & \colhead{$\bigtriangleup E/E$}}
\startdata
4.0 & 33.5  & 27.7 & $5.98\times10^{13}$ & $1.96\times10^{15}$ & 3.04\%\\
6.0 & 36.1  & 32.7 & $7.64\times10^{13}$ & $1.99\times10^{15}$ & 3.84\%\\
8.0 & 40.2  & 36.1 & $8.15\times10^{13}$ & $2.00\times10^{15}$ & 4.08\%\\
10.0 & 43.6 & 37.1 & $9.24\times10^{13}$ & $2.03\times10^{15}$ & 4.55\%
\enddata
\tablecomments{$\Phi_z^c$ represents the catastrophic point; $\bigtriangleup L_c$ is the spatial amplitude of the catastrophe; $\bigtriangleup E$ is the deduced magnetic energy per unit length in $z$-direction within the computational domain; $E$ is the total magnetic energy within the domain just before the catastrophe; $\bigtriangleup E/E$ represents the proportion of the deduced magnetic energy.}
\end{deluxetable}

\begin{deluxetable}{cccccc}
\tablewidth{0pt}
\tablecaption{Parameters of the catastrophes under different $w$ with $d=10.0$ Mm\label{tbl:c}}
\tablehead{ \colhead{$w$(Mm)} & \colhead{$\Phi_z^c$($10^{10}$ Wb)} & \colhead{$\bigtriangleup L_c$(Mm)} & \colhead{$\bigtriangleup E$(J m$^{-1}$)}& \colhead{$E$(J m$^{-1}$)} & \colhead{$\bigtriangleup E/E$}}
\startdata
10.0 & 7.8   & 12.7 & $0.66\times10^{13}$ & $0.32\times10^{15}$ & 2.10\%\\
15.0 & 16.0  & 21.1 & $2.14\times10^{13}$ & $0.58\times10^{15}$ & 3.69\%\\
20.0 & 29.1  & 24.9 & $3.87\times10^{13}$ & $0.95\times10^{15}$ & 4.06\% 
\enddata
\tablecomments{The meanings of the parameters are the same as those in \tbl{tbl:s}.}
\end{deluxetable}

\end{document}